\documentclass[12pt]{JHEP3}

\usepackage{bbm,amsmath,amssymb}
\usepackage{graphicx}
\makeatletter \@addtoreset{equation}{section} \makeatother

\def\ap{{\alpha'}}

\def\2{\frac12}
\def\4{\frac14}
\def\ie{{\it i.e.}~}
\def\eg{{\it e.g.}~}
\def\tx{{\tilde{x}}}
\def\ty{{\tilde{y}}}
\def\tr{{\tilde{r}}}

\newcommand{\be}{\begin{equation}}
\newcommand{\ee}{\end{equation}}
\newcommand{\bea}{\begin{eqnarray}}
\newcommand{\eea}{\end{eqnarray}}
\def\a{\alpha}
\def\b{\beta}
\def\g{\gamma}

\def\m{\mu}
\def\n{\nu}

\author{Massimo Bianchi and Elisa Trevigne \\
\email{Massimo.Bianchi, Elisa.Trevigne@roma2.infn.it}}
 \abstract{ We compute the one-loop  partition function and analyze the
conditions for tadpole cancellation in type I theories
compactified on tori in the presence of internal {\sl oblique}
magnetic fields. We check open - closed string channel duality and
discuss the effect of T-duality. We address the issue of the
quantum consistency of the toroidal model with stabilized moduli
recently proposed by Antoniadis and Maillard (AM). We then pass to
describe the computation of one-loop threshold corrections to the
gauge couplings in models of this kind. Finally we briefly comment
on coupling unification and dilaton stabilization in
phenomenologically more viable models. } \preprint{ROM2F/2005/11
 \\ hep-th/0506080}
\title{Gauge thresholds in the presence of {\sl oblique} magnetic fluxes}

\begin{document}

\section{Introduction and Summary}\label{intro}
>From the very beginning, Type I string theory in the presence of
internal magnetic fields has offered a host of interesting effects
\cite{ft,acny,clny,nestmag,bp,bachas}. From  a theoretical point
of view, such models are governed by exactly solvable conformal
field theories on the worldsheet. The effect of constant abelian
field strengths is reflected in the change of boundary conditions
for the string coordinates. As a result, perturbative analyses are
reliable. From a phenomenologically point of view, magnetized
pan-branes or their T-dual branes at angles are the most promising
candidate to describe semirealistic string vacua that can capture
the essential features of the Standard Model or some of its
supersymmetric and/or grand unified extensions
\cite{branereviews}.

Turning on internal abelian magnetic fluxes reduces the rank of
the CP group to the subgroup  commuting with the $U(1)$
generators. Chiral fermions may arise in the spectrum and the
number of generations, related to the  degeneracy of the Landau
levels, is a topological number that coincides with the top Chern
class of the internal gauge bundle. Since particles interact with
a magnetic field according to their helicity, the degeneracy
between bosons and fermions is in general removed
\cite{bachas,mbys,ads,adds}. However special configurations can
preserve some supersymmetry \cite{angsagrev,gpml}.

 Very recently,  a new mechanism for moduli stabilization has been
proposed \cite{am}  based on the use of {\sl oblique} magnetic
fields on non-factorizable tori. Along this line of investigation,
in \cite{mbet} we have described the effect of arbitrary magnetic
fields on toroidal compactifications of the type I superstring in
various dimensions. In the case of ${\bf T}^6$, one can  attempt
the stabilization of all closed string moduli, except dilaton and
axion, through the introduction of suitable {\sl oblique} choices
of internal abelian magnetic fluxes while preserving a common
${\cal N}=1$ supersymmetry. Unfortunately cancelling all tadpoles
both in the R-R and NS-NS sector seems to be harder to achieve
than originally proposed in \cite{am}\footnote{We thank the
referee for pointing us out possible problems with tadpole
cancellation in the AM model and acknowledge clarifying
discussions on this issue with I. Antoniadis and T. Maillard.}. In
\cite{mbet} we have also identified the tree level gauge couplings
of the surviving Chan-Paton group commuting with the magnetic and
thus anomalous $U(1)$'s \cite{u1anom}.

In the present paper we would like to extend our analysis to
one-loop and compute threshold corrections to the open string
gauge couplings. In principle one can play with the  rationally
quantized values of the internal magnetic fields in order to
adjust the thresholds in closely related phenomenologically viable
models and make contact with low-energy inputs.

The plan of the paper is as follows. In section \ref{partfunct} we
fill in some gaps left open in \cite{mbet} and write down detailed
formulae for the Annulus ${\cal A}$ and M\"obius-strip ${\cal M}$
contributions to the one-loop open string partition function in
the presence of internal {\sl oblique} magnetic
fields\footnote{The torus ${\cal T}$ and Klein-bottle ${\cal K}$
contributions to the unoriented closed string partition function
at one-loop are unaltered, since the terms responsible for lifting
the moduli are of order half-loop (disk).}.  As familiar  from the
analysis of unoriented open strings stretched between branes with
`parallel' magnetic fields, there are several sectors. Neutral
strings connecting branes without magnetic fields preserve ${\cal
N}=4$ supersymmetry and were described long ago
\cite{bps,mbtor,wittor}. Singly charged strings connecting neutral
branes to  magnetized branes can at most preserve ${\cal N}=2$ or
${\cal N}=1$ supersymmetry in $D=4$. Generically supersymmetry is
completely broken in these sectors. When the rank of the magnetic
flux is not maximal, such as in the ${\cal N}=2$ cases, open
strings carry generalized zero modes which are combinations of KK
momenta and windings determined by the orientation of the magnetic
field wrt the fundamental cell of the torus ${\bf T}^6$. When the
rank of the magnetic flux is maximal, such as in ${\cal N}=1$
cases, open strings carry discrete multiplicities determined by
the index $I_{ab}$ of the internal Dirac operator coupled to the
magnetic field. In the unoriented case there are also doubly
charged strings stretched between magnetized branes and their
images under world-sheet parity $\Omega$. Moreover there are
dipole strings having their ends on the same (stack of) branes and
thus preserving ${\cal N}=4$ supersymmetry but carrying `rescaled'
momenta. Finally one has dy-charged strings connecting branes with
different magnetic fluxes, generically {\sl oblique} wrt one
another. As shown in \cite{mbet}, in order to determine the
magnetic shifts one has to diagonalize the orthogonal matrix \be
R_{ab}=R_a^{q_a}R_b^{q_b} \ee where $q_a, q_b= \pm 1$ account for
the (relative) orientation of the two ends and \be R(H) = {1 -
H\over 1+H } \quad , \ee with $H_{\tilde{i}\tilde{j}}(F) =
E_{\tilde{i}}^i E_{\tilde{j}}^j F_{ij}$ the `frame' components of
$F_{ij}$. We will mostly concentrate on supersymmetric
configurations and derive the detailed open string spectrum.
Switching to the transverse closed string channel we check
consistency with the boundary state formalism
\cite{clny,acny,divelic} where magnetic shifts show up as phases
modulating the reflection coefficients.

In section \ref{thresh} we pass to consider the effect of turning
on an abelian magnetic field in two of the four non-compact
directions \eg \be F_{\m\n} = \delta_{[\m}^2 \delta_{\n]}^3 f Q
\ee where $Q$ is one of the generator of the unbroken Chan-Paton
(CP) group. Since the spacetime magnetic 'deformation' is
integrable one can easily write down the relevant contributions:
${\cal A}(f)$ and ${\cal M}(f)$. The closed string spectrum is
unaltered to the order at which we work and plays no role in our
analysis. Selecting the terms quadratic in $f$ (and thus in $Q$)
and subtracting the IR (in the open string channel)
logarithmically divergent terms responsible for their running, we
present general formulae for the one-loop threshold corrections to
the gauge couplings \cite{abd}. After diagonalization of the
magnetic rotation matrices, our formulae look very much the same
as in the case of 'parallel' magnetic fields \cite{paraluestetal}
which in turn show some similarity with standard formulae for
orbifolds \cite{abd,anastas}. We can thus exploit the available
technology in order to write very explicit formulae for the
thresholds arising from both ${\cal N}=2$ and ${\cal N}=1$
sectors\footnote{${\cal N}=4$ sectors, neither contribute to the
(IR) running nor to the thresholds.}. In principle the threshold
corrections under consideration might be completely determined if
the other closed string moduli, except for the overall dilaton
dependence, were fixed, in a supersymmetric fashion, by a proper
choice of internal magnetic fields following the original proposal
of \cite{am}. Unfortunately, we have not been able so far to
achieve this goal in a way consistent with tadpole cancellation in
the absence of orbifolds or lower dimensional $\Omega$-planes.

In section \ref{fine} we conclude with some remarks  on dilaton
stabilization \cite{modstab,fluxgaug} and a preliminary discussion
on the effect of turning on open string Wilson line moduli and
their mixing with closed string moduli. We also pay some attention
to other low-energy couplings most notably Yukawa couplings.

\section{Toroidal compactifications with {\sl oblique}
magnetic fluxes}\label{partfunct}

The perturbative spectrum of unoriented strings\footnote{These are
sometimes referred to as `open descendants', `(un)orientifolds',
type I strings, ... They typically but not necessarily require
open strings for consistency.} is coded in four one-loop
amplitudes \cite{as,gpas,mbas,angsagrev}. Torus ${\cal T}$ and
Klein-bottle ${\cal K}$ represent the contribution of the
unoriented closed strings. Annulus ${\cal A}$ and M\"obius strip
${\cal M}$ represent the contributions of unoriented open strings.
Our aim in this section is to compute the  open string partition
function for toroidal compactifications in the presence of {\sl
oblique} magnetic fluxes.

Toroidal compactifications of type I strings without magnetic
fluxes were studied long ago \cite{bps}. The role of open string
Wilson lines in the `adjoint' breaking of the CP group was
streamlined. Rank reduction due to a quantized NS-NS antisymmetric
tensor background was first pointed out and then further clarified
 in connection with
non-commuting Wilson lines \cite{kaku,mbtor,wittor}, shift
orbifolds \cite{angbab,prator} and exotic $\Omega$-planes
\cite{wittor,dms,angsagrev}.  Special features of rational points
were analyzed. Last but not least, the RR emission vertex in the
asymmetric superghost picture (-1/2,-3/2) was proposed that
involves the RR gauge potential rather than its field
strength\footnote{Though hardly recognized in the overwhelming
literature on D-branes, in retrospect this vertex accounts for
their RR charge and BPS-ness.}.

Turning on magnetic fields does not change the one-loop closed
string amplitudes ${\cal T}$ and ${\cal K}$ and thus the closed
string spectrum to lowest order (sphere), but does affect the open
string spectrum and by open-closed string duality the boundary
reflection coefficients. So far only partition functions for cases
with 'parallel' fluxes have been  explicitly computed, see \eg
\cite{angsagrev,dt,gp,carlonew}. We will momentarily adapt and
extend those results to the case of arbitrary magnetic fluxes.

\subsection{Open string partition function}

Let us divide the full set of branes into various stacks $N_0$,
$N_1$, ... and turn on constant magnetic fields on each stack
except for the first ($a=0$) that we leave unmagnetized. The
resulting gauge group is $SO(N_0)\times U(N_1) \times ...$. As
shown in \cite{am}, the magnetic $U(1)$'s are anomalous and the
corresponding photons become massive by eating R-R axions
associated to internal (1,1)  forms\footnote{This is a rather {\it
petite bouffe} for one of the present authors' standards.}.
Henceforth we will focus on the case of ${\bf T}^6$ for
definiteness and suppress the integration measure $\int dt/t$ as
well as the (regulated) contribution of the zero modes of the four
non-compact bosonic coordinates $V_4/(4 \pi^2 \ap t)^2$.

\subsection{Neutral and Dipole strings}

The annulus contribution ${\cal A}_{00}$ from the completely
neutral strings is the same as for toroidal compactifications
without fluxes \cite{bps} \be {\cal A}_{00} =  {1\over 2} N_0^2
{\cal Q}(0|\tau_A) \sum_{p_{oo}\in \Lambda_{KK}}\exp(2{\pi i \ap
p_{oo}^2 \tau_A}) \ee where $\tau_A = it/2$ and \be {\cal
Q}(\zeta^I |\tau) = {1\over 2} \sum_{\a,\b} c_{\a\b}
{\theta\left[^\a_\b\right](0|\tau) \over \eta^{3}(\tau)}
\prod_{I=1}^3{i \theta\left[^\a_\b\right](\zeta^I| \tau) \over
\theta_1(\zeta^I|\tau)} \quad , \ee with $c_{\a\b} = \exp[2\pi
i(\alpha+\beta)]$  implementing the GSO projection. As indicated,
only KK momenta $p^{\hat{i}}_{oo}= m^i E^{\hat{i}}_{i}$, with $m^i
\in Z$ (for singly wrapped branes) and $E^{\hat{i}}_{i}$ the
inverse 6-bein, are allowed. For simplicity, we have set the
quantized NS-NS antisymmetric tensor $B_{ij}$ and the open string
Wilson lines $A_i^a$ to zero. We postpone a brief discussion on
their effects to the concluding remarks.

The M\"obius-strip $\Omega$ projection in this sector reads \be
{\cal M}_{00} = -{1\over 2} N_0 {\cal Q}(0|\tau_M )
\sum_{p_{oo}\in \Lambda_{KK}} \exp({2 \pi i\ap p_{oo}^2 \tau_A})
\ee where $\tau_M = \tau_A +1/2$ and, again, only KK momenta are
allowed.

Neutral dipole strings starting and ending on the same stack
$a\neq 0$ of branes suffer no magnetic mode shifts and contribute
\be {\cal A}_{a\bar{a}} = N_a \bar{N}_a {\cal Q}(0|\tau_A)
\sum_{p_{a\bar{a}}\in  \Lambda_{a\bar{a}}} \exp({2 \pi i \ap
\tau_A p_{a\bar{a}}^2} )
 \ee where the lattice
sum is over generalized momenta $p_{a \bar{a}}$, satisfying $p_L =
R_a p_R$, which generalizes the condition $p_L = p_R$ valid for
truly neutral strings, discussed above.

\subsection{Singly and doubly charged strings}

Singly charged strings, connecting unmagnetized branes to
magnetized ones, are easy to analyze too. The magnetic shifts read
\be \epsilon^{I}_{oa} = {1\over \pi } \arctan(q_a h^I_a)\quad ,
\ee where $h^I_a$ with $I=1,2,3$ are the skew eigenvalues of
$H^a_{\hat{i}\hat{j}} = E^{i}_{\hat{i}} E^{j}_{\hat{j}} F^a_{ij}$
(frame components!), and turn the supersymmetric `character'
${\cal Q}(0|\tau_A)$ into ${\cal Q}(\epsilon^I_{oa}\tau_A
|\tau_A)$.

The overall multiplicity, related to the degeneracy of the Landau
levels, is given by\footnote{We assume $I_{oa}$ to be positive. A
negative $I_{oa}$ would imply the presence of massless fermions of
opposite chirality in the open string spectrum.} \be I_{oa} =
|W_a| V({\bf T}^6) \prod_I q_a h^I_a = \prod_I q_a m_a^I \ee where
$V({\bf T}^6)$ is the `volume' of ${\bf T}^6$ in units of
$4\pi^2\ap$, \be |W_a|= \det\left(\partial X^i\over
\partial\sigma^\alpha\right) = \prod_I n^I_a \ee is the integer
wrapping number, and $m^I_a$  are the integer magnetic monopole
numbers. Dirac quantization indeed constraints the skew
eigenvalues of $(2\pi\ap) F^a_{ij}$ (adimensional!) to be given by
$f^I_a = m_a^I/n^I_a$. If ${\bf T}^6 = \prod_I {\bf T}^2_{(I)}$
then $V({\bf T}^6) = \prod_I V_I$, with $V_I$ the `volume' of
${\bf T}^2_{(I)}$ and $f^I_a = V_I h^I_a$. In any case, it is easy
to prove that \be I_{oa} = |W_a| V({\bf T}^6) \prod_I {\sin(\pi
\epsilon^I_{oa})\over \cos(\pi \epsilon^I_{oa})} = \prod_I
\sin(\pi \epsilon^I_{oa}) \sqrt{\det({\cal G}_a + {\cal F}_a)}
\quad ,\ee where ${\cal G}_a$ and ${\cal F}_a$ are the induced
worldvolume metric and field strength. This has a clear
interpretation in the transverse channel, where it exposes the
Born-Infeld (BI) action \cite{leigh,mss}.  The extra product of
sinus turns out to cancel a similar factor coming from the
$\theta_1$'s in the denominator.

If one or more of the $h^I_a $ are zero, \ie $H^a$ has not maximal
rank, the index vanishes signalling the presence of invariant
subtori\footnote{The label $u$ indeed stands for `unmagnetized'.}
${\bf T}^2_u$, skewly embedded in ${\bf T}^6$. The `unmagnetized'
directions are those fixed under $R_a$ and along them the open
 string can carry generalized momenta simultaneously satisfying
$p_L = R_a^{q_a}
p_R$, at the charged end, and $p_L = p_R$, at the neutral end $a=0$.
Compatibility of the two conditions follows from $\det(R_a^{q_a} - 1) =0$.

Generically there are tachyons in these sectors since all susy
tend to be broken by the presence of the magnetic flux
\cite{bachas,mbys,ads}. However when $\sum_I (\pm)_I
\epsilon^{I}_{0a} = 0$ for some choice of signs the magnetic
rotation matrix $R_a^{q_a}$ belongs to an $SU(3)$ subgroup of
$SO(6)$ and the sector is at least ${\cal N} = 1$ supersymmetric
\cite{bala,berk,raba}. If  moreover one of the three mode shifts
is zero, let us say $\epsilon^{u}_{0a} = 0$ for $u=3$, then
$\epsilon^{1}_{0a} = \pm \epsilon^{2}_{0a}$ and the sector
preserves ${\cal N} = 2$ susy. Thanks to some Jacobi theta
function identities the annulus amplitudes vanish in both cases
and  read \be {\cal A}_{oa}^{{\cal N} = 1} = N_0 N_a I_{oa} {\cal
Q}(\epsilon^I_{oa}\tau_A|\tau_A) \ee or \be {\cal A}_{oa}^{{\cal
N} = 2} = N_0 N_a \Lambda^u_{oa} I^\perp_{oa} {\cal
Q}(\epsilon^I_{oa}\tau_A|\tau_A) \quad , \ee where
$\Lambda^u_{oa}$ denotes the lattice sum in the unmagnetized
complex direction, and \be I^\perp_{oa} = \prod_{I\neq u} q_a
m_a^I \ee is the reduced index that counts the degenaracy of the
Landau levels in the (four) transverse magnetized directions.

As mentioned above, the lattice $\Lambda^u_{ao}$ consists in those
generalized momenta that satisfy $p_L = R_a^{q_a} p_R$ and
$p_L=p_R$. For generic choices of magnetic fluxes and thus $R_a$
the resulting momenta are neither pure KK nor pure windings but
rather mixtures of the two. The `unmagnetized' directions satisfy
$R_a u = u$ or, equivalently,  $H_a u =0$.

Doubly charged strings connecting magnetic branes with their
$\Omega$ images carry CP multiplicity $N_a^2$ or $\bar{N}_a^2$,
depending on the sign of $q_a= q_b$, suffer doubled magnetic mode
shifts $\epsilon^I_{aa} = 2\epsilon^I_{0a}$ and appear with
rescaled degeneracy of the Landau levels \be I_{aa} = \prod_I 2
q_a m_a^I= 8 I_{oa} \quad , \quad I^\perp_{aa} = \prod_{I\neq u} 2
q_a m_a^I= 4 I^\perp_{oa} \quad . \ee Moreover, this sector
receives a M\"obius strip contribution.

For supersymmetric configurations, one has \be {\cal
A}_{aa}^{{\cal N} = 1} = {N^2_a \over 2} 8 I_{oa} {\cal
Q}(2\epsilon^I_{0a}\tau_A|\tau_A) \ee or \be {\cal A}_{aa}^{{\cal
N} = 2} = {N^2_a \over 2} 4 I^\perp_{oa} \Lambda^u_{oa}(\tau_A)
{\cal Q}(2\epsilon^I_{oa}\tau_A|\tau_A) \ee where as indicated
$\Lambda^u_{aa} =\Lambda^u_{oa}$ is the same as in the singly
charged sector\footnote{When $\epsilon^I_{oa}=1/2$ for some $I$ a
separate treatment is required.}.  The M\"obius strip reads \be
{\cal M}_{aa}^{{\cal N} = 1} = {N_a\over 2} \hat{I}_{aa} \hat{\cal
Q}(2\epsilon^I_{0a}\tau_A |\tau_M) \ee or \be {\cal M}_{aa}^{{\cal
N}= 2} = {N_a\over 2} \hat{I}^\perp_{a} \hat\Lambda^u_{oa}(\tau_A)
\hat{\cal Q}(2\epsilon^I_{oa}\tau_A|\tau_M) \ee where {\it a
priori} $- 8 I_{oa}\le \hat{I}_{aa}\le +8 I_{oa}$ and
 $- 4 I_{oa}\le \hat{I}^\perp_{aa}\le +4 I_{oa}$, both with jumps of 2 units,
allow for all possible (anti)symmetrizations under $\Omega$
\cite{angsagrev}. Although in the simple toroidal models we
explicitly consider $\hat{I}_{aa}= {I}_{aa} = 8 I_{oa}$ and
$\hat{I}^\perp_{aa}= {I}^\perp_{a}= 4 I_{oa}$, turning on a non
vanishing $B$ and/or (discrete) Wilson lines may change the
situation \cite{bps,mbas,angsagrev}.

\subsection{Dy-charged strings}

We are now ready to discuss the last and most subtle case of open
strings strecthed between branes with {\sl oblique} magnetic
fields. As obvious we will recover the simpler case of parallel
magnetic fields as a limit. As shown in \cite{mbet}, in order to
compute the magnetic shifts one has to diagonalize the orthogonal
matrix $R_{ab} = R_a^{q_a} R_b^{q_b}$ (in the frame basis!). For
${\bf T}^6$, the eigenvalues come in (three) complex conjugate
pairs $\rho^{I\pm}_{ab} = \exp(\pm i 2 \beta^I_{ab})$.  The
magnetic shifts are then given in general by \be \epsilon^I_{ab} =
{2\over 2\pi}\beta^I_{ab} \quad . \ee If $R_a R_b = R_b R_a$, the
simpler abelian composition rule $\epsilon^I_{ab} =
\epsilon^I_{ao} \pm \epsilon^I_{ob}$ applies. Only the annulus
contributes to these sectors and, depending on the amount of
supersymmetry preserved, reads\footnote{We try to consistently use
$F$ to denote the 2-form in the coordinate basis and $H$ to denote
the antisymmetric matrix in the frame basis.} \be {\cal
A}_{ab}^{{\cal N} = 1} = N_a \bar{N}_b I_{ab} {\cal
Q}(\epsilon^I_{ab}\tau_A |\tau_A) \ee for ${\cal N} = 1$ sectors,
where \be \label{chern3} I_{ab} = C_3({\cal E}_a^{q_a} \otimes
{\cal E}_b^{q_b}) = { W_a W_b\over 3! (2\pi)^3} \int_{{\bf T}^6}
(q_a F_a + q_b F_b)^3 \quad , \ee and \be {\cal A}_{ab}^{{\cal N}
= 2} = N_a \bar{N}_b I^\perp_{ab} \Lambda^u_{ab}(it) {\cal
Q}(\epsilon^I_{oa}\tau_A |\tau_A) \ee for ${\cal N} = 2$ sectors,
where \be \label{chern2} I^\perp_{ab} = C^\perp_2({\cal E}_a^{q_a}
\otimes {\cal E}_b^{q_b}) = { W_a W_b\over 2! (2\pi)^2} \int_{{\bf
T}^4_\perp} (q_a F_a + q_b F_b)^2 \quad , \ee with ${\bf
T}^4_\perp$ denoting the effectively magnetized subtorus,
comprising the coordinates for which $(q_a F_a + q_b F_b)_{ij} X^j
\neq 0$. Along the complementary unmagnetized torus ${\bf T}^2_u$,
open strings carry zero modes that contribute to the lattice
 $\Lambda^u_{ab}$. This consists in those generalized
momenta that satisfy $p_L = R_a^{q_a} p_R$ and $p_L=
R_b^{-q_b}p_R$,  which are compatible with one another since
$R_{ab} = R_a^{q_a}R_b^{q_b}$ has unit eigenvalues as a
consequence of $(q_aH_a + q_bH_b)$ having zero eigenvalues. For
generic choices of magnetic fluxes and thus $R_a$ and $R_b$ the
resulting momenta are neither pure KK nor pure windings but rather
mixtures of the two. At first sight, there seems to be some
ambiguity in the definition of the `unmagnetized' directions in
these sectors. Indeed $R_a R_b u = u$ implies $R_b R_a v = v$ for
$v=R_b u$ and also $(H_a + H_b)z = 0$ for $z = (1 + H_b)^{-1} u$.
However the three possible choices ($u$, $v$ or $z$) are
equivalent in that they yield the same results for the masses and
multiplicities of the open string states. We will check this
statement by means of T-duality in section \ref{Tdual}.

The above expressions clearly encompass
${\cal A}_{\bar{a}b} = {\cal A}_{a\bar{b}}$ and ${\cal
A}_{\bar{a}\bar{b}} = {\cal A}_{ab}$, upon properly choosing
the signs $q_a$ and $q_b$.

Eqs. (\ref{chern3}) and (\ref{chern2}) indicate that the
degeneracy of the Landau levels in each sector is given by the
relevant Chern class of the internal tensor gauge bundle, which in
turn coincides with the index of the Dirac operator coupled to the
combined magnetic fields. For the purpose of checking consistency
with the transverse closed string channel and accounting for the
emergence of the BI and Wess-Zumino (WZ) terms, it is crucial to
observe that \bea
 I_{ab} & = &  V({\bf T}^6) \prod_{I}n_a^I n_b^I
\frac{\sin(\pi \epsilon^I_{ab})}{ \cos(\pi\epsilon^I_a)
\cos(\pi\epsilon^I_b)} \nonumber\\
& = & V({\bf T}^6) W_a W_b \sqrt{\det(1
+q_aH_a)} \sqrt{\det(1 +q_bH_b)} \prod_{I=1}^3 { \sin(\pi
\epsilon^I_{ab})} \label{indtobi}
\eea
and
\bea
 I^\perp_{ab}
\Lambda^{(u)}_{ab} & = & V({\bf T}^6) \prod_{I\neq u} n_a^I n_b^I
\frac{\sin(\pi \epsilon^I_{ab})}{ \cos(\pi\epsilon^I_a)
\cos(\pi\epsilon^I_b)}\nonumber\\
& = &  V({\bf T}^6) W_a W_b  \sqrt{\det(1 +q_aH_a)} \sqrt{\det(1
+q_bH_b)} \prod_{I\neq u} \sin(\pi\epsilon^I_{ab})
\tilde\Lambda^{(u)}_{ab} \quad . \label{indperptobi} \eea
 To this
end, using $h_{a}^I = \tan(\pi\epsilon^I_{a})$ and elementary
trigonometry, one first expresses the BI action in the form \be
W_a \sqrt{\det (1 +q_aH_a)} = \prod_I n_a^I\sqrt{(1 + (h_a^I)^2)}
= \prod_I {n_a^I \over \cos( \pi\epsilon^I_{a})} \quad . \ee

In the case of parallel magnetic fluxes, one then has \be
\sin(\pi\epsilon^I_{ab}) = \sin(\pi\epsilon^I_{a} +
\pi\epsilon^I_{b}) =  {(q_a h_a^I + q_b h_b^I) \over
\sqrt{1+(h_a^I)^2} \sqrt{1+(h_b^I)^2}} \ee and the denominator can
be used to cancel the BI factors and to get precisely
(\ref{indtobi}).

In the case of arbitrary magnetic fluxes, one has to  work a
little harder \cite{mbet}. The product of sines $\prod_I
\sin(\pi\epsilon^I_{ab})$ can be related to the characteristic
polynomial of $R_{ab}$, $P(\lambda) = \det(R_{ab}-\lambda I)$,
with $\lambda = 1$\footnote{Some of its remarkable properties have
been discussed in \cite{lars}.}. Plugging in (\ref{indtobi}) the
trigonometric formula \be\label{sin} \prod_{I\neq u} \sin(\pi
\epsilon^I_{ab}) = \sqrt{{\det^\prime} \left({R_{ab} - 1\over 2}
\right)} = {\sqrt{\det^\prime(q_aH_a + q_b H_b)} \over
\sqrt{\det(1 + q_aH_a)} \sqrt{\det(1 + q_bH_b)}} \quad , \ee
proven in the Appendix, the BI terms in the denominator cancel and
the index reads \bea I_{ab} & = & V({\bf T}^6)\sqrt{\det(1+q_a
H_a)}\sqrt{\det(1+q_b H_b)} \sqrt{\frac{1}{2^6}P(1)} \nonumber\\
& = & V({\bf
T}^6)
\sqrt{\det(q_aH_a + q_b H_b)} =  V({\bf T}^6){\rm Pfaff}(q_aH_a + q_b H_b)
\nonumber\\
& = & \frac{1}{2\cdot  3!}\int_{{\bf T}^6} (q_aF_a+q_b F_b)^3
 \eea
up to signs.

When ${\rm Pfaff}(q_aH_a + q_b H_b) = 0$ \ie when $\det(q_aH_a +
q_b H_b) = 0$  then $\det(R_{a}^{q_a} R_{b}^{q_b} - 1) = 0$ one
has unmagnetized directions along which open strings carry zero
modes, \ie mixtures of KK momenta and windings. We have already
observed that the various kernels are isomorphic. One eventually
finds
\bea I^\perp_{ab} & = & V({\bf T}_\perp^4)\sqrt{\det(1+q_a
H_a)^\perp}\sqrt{\det(1+q_b H_b)^\perp}
\sqrt{\frac{1}{2^4}P^\prime(1)} \nonumber\\
& = & V({\bf T}_\perp^4)
\sqrt{\det(q_aH_a + q_b H_b)^\perp}
 =  V({\bf T}_\perp^4){\rm Pfaff}(q_aH_a + q_b H_b)^\perp \nonumber\\
& = & \frac{1}{2\cdot  2!}\int_{{\bf T}_\perp^4} (q_aF_a+q_b
F_b)^2 \eea as expected.

\section{Channel duality and tadpoles}

In order to check the validity of our derivation of the open
string spectrum, encoded in the direct (loop) open string channel,
we would now like to compute the resulting transverse (tree
level) closed string channel. For consistency one expects to find a
boundary-to-boundary amplitude of the form \be \tilde{\cal A} =
\sum_{a,b} N_a N_b \langle B_a|\exp(-\pi \ell {\cal H}_{cl})
|B_b\rangle  \quad , \ee
where the presence of ${\cal H}_{cl}$ means that
only states in the closed string spectrum are allowed to be exchanged.

The superstring boundary state $ |B(F)\rangle $ in the presence of
an  arbitrary (electro-)magnetic field was constructed long ago
\cite{clny,acny} and reconsidered more recently \cite{divelic}. It
consists of various ingredients and obviously depends on the
choice of boundary conditions for the worldsheet supercurrent, \ie
for the worldsheet fermions and superghosts. The ghost
contribution  is independent from the magnetic flux and we will
not display it for simplicity. Indeed, since all electric
components vanish in our case, $F_{io}=0$, we can choose a
light-cone gauge and work with the eight transverse coordinates
only, $i,j=2,...9$, and forget about (super)ghosts altogether.

The contribution of the bosonic coordinates \be |B_a\rangle^{(X)}
= \sqrt{\det ({\cal G}_a + {\cal F}_a)} \exp (- \sum_{n>o}
\tilde{a}^i_{-n} R_{ij}(F_a) a^j_{-n}) |O_a\rangle \ee Here we are
back to the coordinate basis, where $R_{ij}$ is not an orthogonal
matrix! One switches from one to the other by means of
$E_{\hat{i}}^i$ ad its inverse $E^{\hat{i}}_i$. Taking into
account the obvious generalization  associated to multiple
wrapping and the presence of a (flat) non-trivial induced metric.
Notice the presence of the BI action that generalizes the overall
volume contribution of the CM position when $F_a\neq 0$. The
bosonic zero-mode contribution is implicit in $|O\rangle_a $ and
deserves a special treatment. It consists in a sum over all $p_L =
- R_a p_R$. In compact  cases\footnote{For non-compact directions
$p_L = - p_R$ and, even in the presence of magnetic fields, this
results in the familiar Neumann condition of 'no momentum flow'
through the boundary. This observation turns out to be relevant
for our later purposes of computing thresholds.}, this results in
an infinite but discrete number of choices, \eg windings for
$F_a=0$ or generalization thereof for  $F_a\neq 0$.

The contribution of the fermionic coordinates is notoriously much
subtler. In the NS-NS sector, there are no fermionic zero-modes,
since the modes are half-integers and one has \be |B_a, \eta
\rangle_{NS-NS}^{(\psi)} = \exp (-i\eta \sum_{n\ge 1/2}
\tilde{\psi}^i_{-n} R_{ij}(F_a) \psi^j_{-n}) |\eta\rangle \ee
where the $\eta=\pm$ stands for possible GSO  projections and the
light-cone gauge roughly speaking corresponds to the choice of the
canonical (left-right symmetric) superghost picture
$q=\tilde{q}=-1$.

In the R-R sector, fermions admit zero-modes, whose  contribution
replaces the BI action with the WZ coupling \be |B_a, \eta
\rangle_{R-R}^{(\psi)} = {1 \over \sqrt{\det ({\cal G}_a + {\cal
F}_a)}} \exp (i\eta \sum_{n>0} \tilde{\psi}^i_{-n} R_{ij}(F_a)
\psi^j_{-n}) |O_a, \eta\rangle \ee where \be |O_a, \eta\rangle =
{U}_{A\tilde{B}}(F_a) |A,\tilde{B}\rangle \ee with \be
{U}_{A\tilde{B}}(F_a) = \left[{\rm AExp} \left(- {\frac{1}{2}}
F^a_{ij}\Gamma^{ij}\right)\right]_{A\tilde{B}}\quad , \ee where
the notation ${\rm AExp}$ implies that one has to antisymmetrize
the vector indices of the $\Gamma$ matrices \ie \be {\rm AExp}
\left(-{\frac{1}{2}} F^a_{ij}\Gamma^{ij}\right) = 1 - {1 \over 2}
F^a_{ij}\Gamma^{ij}+ {1 \over 8}
F^a_{ij}F^a_{kl}\Gamma^{[ij}\Gamma^{kl]} + ... \quad . \ee

The full boundary state\footnote{A similar  analysis allows one to
construct crosscap states. We refrain from doing so here since
there are no issues at stake for the Klein bottle $\widetilde{\cal
K}$ or M\"obius strip $\widetilde{\cal M}$.}
 then reads \be |B\rangle = {1\over 2}
\sum_a N_a( |B_a,+\rangle_{NS-NS}-|B_a,-\rangle_{NS-NS} +
|B_a,+\rangle_{R-R}+|B_a,-\rangle_{R-R}) \quad .\ee

Let us now consider for definiteness the amplitude \be \tilde{\cal
A}_{ab} = \langle B_a|\exp(-\pi \ell {\cal H}_{cl}) |B_b\rangle
\quad . \ee for $a\neq b$ with $[R_a, R_b]\neq 0$. Since ${\cal
H}_{cl}= L_o + \tilde{L}_o - c/12$ is a (transverse) Lorentz
scalar, $L_o = \ap p_L^2/4 + \sum_{n>o} [n a_{-n} a_{n} +
\psi_{-n}\psi_{n}]$ one can perform a simultaneous rotation of all
$a^i_n$'s (both annihilation $a^i_{n>o}$ and creation $a^i_{-n<o}$
modes) by say $R_b$: $\hat{a}^i_n = R_b^i{}_j a^j_n$ that leaves
$L_o$ invariant and preserves the canonical commutation rules. The
net result is to transfer the effect of the rotation on the other
boundary state $|B\rangle_a$ that, once written in terms of
$\hat{a}^i_n$ and $\tilde{a}^i_n$, depends on the combined
rotation $R_{ab}= R_a R_b^{-1}$. Thence everything, except for the
zero-modes and overall BI or WZ actions, goes through in the same
way as for closed string bouncing between an unmagnetized brane
and a magnetized one. In particular  mode shifts in the direct
channel give rise to phases in the transverse channel. Relying on
the expressions for $I_{ab}$ and $I^\perp_{ab}$ that have been
derived at the end of the previous (sub)section, one can also
reproduce the expected BI action or WZ couplings.

\subsection{UV Divergences}

We are thus ready to address the question of UV divergences in the
presence of {\sl oblique} fluxes and their cancellation. As it is
well known, they are associated to diagrams (tadpoles) of massless
particles (dis)appearing from (into) the vacuum
\cite{mbas8192,mrd}. In particular tadpoles of RR massless fields
belonging to closed sectors with non-vanishing Witten index are
responsible for chiral anomalies in the low energy effective
theory  \cite{pcai,anomtad}. On  the other, hand NS-NS tadpoles
are less dangerous in principle \footnote{For a derivation of the
dilaton tadpole in the type I superstring see \cite{ohta}}. They
{\it simply} signal the instability of the chosen vacuum
configuration \cite{fs,dnpstad}.

Since R-R and NS-NS fields couple to pan-branes according to the
WZ and BI actions \cite{clny,acny}, respectively,  turning on
internal open string fluxes induces lower dimensional  R-R charges
and NS-NS tensions \cite{braninbran}. Quite remarkably, some of
these can be negative for special choices of fluxes on ${\bf T}^6$
preserving at most ${\cal N}=1$ supersymmetry and correspond to
stable bound states not at threshold \cite{marche}. As a
consequence, one may try to satisfy the consistency constraints
without adding  lower dimensional D-branes and/or $\Omega$-planes
\cite{dt,am}. For ${\bf T}^4$ and/or for ${\cal N}=2$
supersymmetric configurations on ${\bf T}^6$, instead, BPS bound
states are necessarily at threshold. Yet, even in these cases, one
can play with the non-polynomial BI action and derive
supersymmetric configurations associated to non-linear instantons
\cite{mmms,raba,mbet,dt}.

In order to expose potential massless tadpoles at  genus 1/2 from
the disk and projective plane, we start by performing an S modular
trasformation $\tau_A = \,\frac{i t}{2}\, \rightarrow
-\frac{1}{\tau}_A=i \ell\;$ on the Annulus amplitude ${\cal A}$
and get
 \bea
\tilde{\cal A}_{ab} & = & {N_a N_b |W_a||W_b| \over 2^{5}} \int
d\ell \sqrt{\det(G_{ab})}\widetilde{\Lambda}_{ab}(i\ell)
\sqrt{\det(1 +q_aH_a)} \sqrt{\det(1 +q_bH_b)}
\nonumber\\
&  & \sum_{\a\b} \frac{c_{\a\b}}{2}
\left(\theta\left[^\a_\b\right](0 |i\ell) \over
\eta^3(i\ell)\right)^{1+u}\;
\prod_{I=1}^{3-u}{\theta\left[^\a_\b\right] (i\epsilon^I_{ab}|
i\ell) \over
\theta\left[^\frac{1}{2}_\frac{1}{2}\right](i\epsilon^I_{ab}|
i\ell)}  {2\sin(\pi\epsilon^I_{ab})} \quad ,\eea where we use the
modular properties of Jacobi theta functions and open/closed
duality for the index. The zero mode contribution can be dealt
with by means of a Poisson resummation \be
\widetilde\Lambda_{ab}(i\ell)= \sum_{w_{ab}}\exp\left(-\frac{2 \pi
\ell\ w_{ab}^{i}G_{ij} \, w_{ab}^{j}}{4\ap}\right)
 \ee
Modular transformation properties of $\theta$ and other functions
are listed in an Appendix.

Similarly, for the M\"obius strip ${\cal M}$, the relevant modular
trasformation is $P=TST^2S$, that acts on the modular parameter
according to $\tau_M = \,\frac{i t +1}{2}\, \rightarrow
\frac{1}{2}+\frac{i}{2 t}= \frac{1}{2} + i \ell\;$ so that $\,\ell
= 2t $, and yields \bea \widetilde{\cal M}_{a} & = & - 2 N_a |W_a|
\int d\ell \sqrt{\det(G_{aa})}\widetilde{\Lambda}_{aa}(i\ell)
\sum_{\a\b} \frac{c_{\a\b}}{2} \left(\theta\left[^\a_\b\right](0
|i\ell) \over \eta^3(i\ell)\right)^{1+u}\;
\prod_{I=1}^{3-u}{\theta\left[^\a_\b\right](i\epsilon^I_{aa}/2|
i\ell) \over
\theta\left[^\frac{1}{2}_\frac{1}{2}\right](i\epsilon^I_{aa}/2| i\ell)}\nonumber\\
&  & \sqrt{\det(1
+q_aH_a)}\prod_{I=1}^{3-u}{2\sin(\pi\epsilon^I_{aa}/2)}.
 \eea
At this order the unoriented closed string spectrum is unaffected
by the internal magnetic field and the Klein bottle amplitude
${\cal K}$ gives rise to \be \widetilde{\cal K} = 2^5\int d\ell
\sqrt{\det(G_{6})}\Lambda_{6}(i\ell) \sum_{\a\b}
\frac{c_{\a\b}}{2}\left(\theta\left[^\a_\b\right](0 |i\ell) \over
\eta^3(i\ell)\right)^4  \quad ,\ee after an S modular
transformation.

In the RR sector, the (closed string) IR limit  $ \ell \rightarrow
\infty$ is dominated by the exchange of massless states and yields
\bea\label{Ampl} \widetilde{\cal A}^{R-R}_{m=0} & \sim & 2^{-5}
8_s \sum_{a b}\sum_{q_a q_b} N_{a}^{q_a}W_a
N_{b}^{q_b}W_b\sqrt{\det(1 +q_a H_a)}\sqrt{\det(1 +q_b H_b)}
\prod_{I=1}^{3}{\cos(\pi\epsilon^I_{ab})}\nonumber\\
\widetilde{\cal M}^{R-R}_{m=0} & \sim & -2 8_s \sum_{a q_a}  \, N_a^{q_a} W_a \nonumber\\
\widetilde{\cal K}^{R-R}_{m=0} & \sim & 2^5 \, 8_s \eea The
deceiving simplicity of $\widetilde{\cal M}_{m=0}$ is to be
ascribed to the often used identity \be \sqrt{\det(1 + q_a H_a)} =
1/ \prod_{I=1}^{3}{\cos(\pi\epsilon^I_{a0})} = 1/
\prod_{I=1}^{3}{\cos(\pi\epsilon^I_{aa}/2)} \ee

At this point it is useful to take advantage of the spinorial
representation of the rotation matrix $R(F)$, introduced above,
\be U(R) = \frac{1}{\sqrt{\det(1 + q H)}}{\rm AExp}
\left(-\frac{q}{2} \g^{\hat{i}}\g^{\hat{j}}
H_{\hat{i}\hat{j}}\right ) \ee in order to recognize that
$$
\prod_{I=1}^{3}{2\cos(\pi\epsilon^I_{ab})}= Tr_s[U(R_{ab})]=
Tr_s[U(R_a)U(R_b)]
$$
or, equivalently, \be \prod_{I} 2\cos(\pi \epsilon^I_{ab}) =
\sqrt{\det\left(R_{ab} + 1 \right)} =  {2^3 \sqrt{\det(1 + q_aH_a
q_b H_b)} \over \sqrt{\det(1 + q_a H_a)} \sqrt{\det(1 +
q_bH_b)}}\ee

In this case after a sum over the  possible orientations,
 $q=\pm 1$, only wedge products with even numbers of $H$ survive.
Tracing over the spinor indices of $\gamma$ matrices, the above
expression factorizes in a sum of squares that are easily
interpreted in terms of the total R-R charge of D9-branes,
$\Omega$9-plane and of the individual R-R charges of the lower
dimensional objects induced by the fluxes.  Observing that the
series in ${\rm AExp}$ actually truncates at order $d/2=3$ in our
case and that
$$
{\cal Q}_a^{\hat{i}\hat{j}} = {1\over 2^3}
\epsilon^{\hat{i}\hat{j}\hat{i}_1\hat{j}_1\hat{i}_2\hat{j}_2}
H^a_{\hat{i}_1\hat{j}_1}H^a_{\hat{i}_2\hat{j}_2}
$$
accounts for the induced D5-brane charge  of the magnetized
D9-branes, one eventually finds the complete R-R tadpole condition
\bea &  & \widetilde{\cal A}^{R-R}_{m=0}+\widetilde{\cal
K}^{R-R}_{m=0}+\widetilde{\cal M}^{R-R}_{m=0} = \left(\sum_{a} 2
N_a W_a -\,32\right)^2 + \sum_{\hat{i}\,\hat{j}}\sum_{a\,b} N_a
W_a
N_b W_b {\cal Q}_a^{\hat{i}\hat{j}}{\cal Q}_b^{\hat{i}\hat{j}}\nonumber\\
& = & \left(\sum_{a} 2 N_a W_a - \,32\right)^2 +
\sum_{\hat{i}\,\hat{j}}\left ( \sum_{a} N_a W_a {\cal
Q}_a^{\hat{i}\hat{j}} \right )^2 = 0 \eea This consistency
condition, derived here  using CFT techniques \ie channel duality,
coincides with the one based on the analysis of the BI and WZ
actions \cite{am} or on anomaly cancellation
arguments\cite{anomtad}. Unfortunately, due to subtleties with the
choice of the wrapping numbers, the AM model does not satisfy
these consistency requirement even if the spectrum is not chiral.
We have not been able so far to find consistent variants of the AM
model, although to the best of our knowledge no 'no-go' theorem
prevents their existence. One possible way out would be to include
magnetized ${\bar D}9$-branes, preserving the same ${\cal N} =1$
susy and corresponding to $W_a <0$.

By similar means one can study the NS-NS sector. Taking the limit
$\ell \rightarrow \infty$ one finds \bea\label{Ampl}
\widetilde{\cal A}^{NS-NS}_{m=0} & \sim & 2^{-5}\sum_{a
b}\sum_{q_a q_b} N_{a}^{q_a}|W_a| N_{b}^{q_b}|W_b|\sqrt{\det(1
+q_a H_a)}\sqrt{\det(1 +q_b H_b)}\nonumber\\
& & \times [2  + \sum_{I=1}^{3} 2\cos(2\pi\epsilon^I_{ab}) ]\nonumber\\
\widetilde{\cal M}^{NS-NS}_{m=0} & \sim & -2  \sum_{a q_a} 8 \,
N_a^{q_a} |W_a| \sqrt{\det(1 +q_a H_a)}
[2  + \sum_{I=1}^{3} 2\cos(2\pi\epsilon^I_{ao}) ]\nonumber\\
\widetilde{\cal K}^{NS-NS}_{m=0} & \sim & 2^5 \, 8_v \quad .\eea
Using \be 2 \sum_{I} \cos(2\pi \epsilon^I_{ab}) =  \sum_{I}
(e^{2i\pi \epsilon^I_{ab}}+ e^{-2i\pi \epsilon^I_{ab}}) =
Tr_v(R_{ab}) = Tr_v(R_{a}R_{b}) \quad ,\ee proven in the appendix
yields the complete massless NS-NS tadpole condition \bea &  &
\widetilde{\cal A}^{NS-NS}_{m=0}+\widetilde{\cal
K}^{NS-NS}_{m=0}+\widetilde{\cal M}^{NS-NS}_{m=0}
\nonumber\\
& = & \left( \sum_{a} 2 N_a |W_a| \sqrt{\det(1 +q_a H_a)} -
\,32\right)^2 + \sum_{\hat{i}\, \hat{j}}\sum_{a\,b} N_a^{q_a}
|W_a|
N_b |W_b| {\cal T}_a^{\hat{i} \hat{j}}{\cal T}_b^{\hat{i} \hat{j}}\nonumber\\
& = & \left(\sum_{a } N_a  |W_a| \sqrt{\det(1 +q_a H_a)} - \,32
\right)^2 + \sum_{\hat{i}\,\hat{j}}\left ( \sum_{a } N_a |W_a|
{\cal T}_a^{\hat{i}\hat{j}} \right )^2 = 0 \eea The overall
tension of the bound state  of magnetized D9-branes is positive,
being the positive branch of a square root. As a result, the
vanishing of the dilaton tadpole, despite the presence of the
negative contribution from the tension of the $\Omega$9-plane,
seems hard to achieve with non-trivial fluxes compatibly with
RR-tadpole cancellation. The remaining massless tadpoles indicate
the presence of induced lower dimensional tensions of both signs,
which in turn are derivatives of the potential generated by the BI
couplings \be {\cal T}^a_{\hat{i}\hat{j}} = E_{\hat{i}}^i
E_{\hat{j}}^j {\partial{\cal V}_a \over  \partial G^{ij}}\quad .
\ee Indeed it is easy to prove that
\be E_{\hat{i}}^i
E_{\hat{j}}^j {\partial{\cal V}_a \over
\partial G^{ij}} = {1\over 2} (R^a_{\hat{i}\hat{j}} +
R^a_{\hat{j}\hat{i}}) = {1\over 2} \sum_{q_a=\pm 1}
R^{a\,q_a}_{\hat{i}\hat{j}} \quad .\ee

Let us stress once again that configurations of this kind  are not
bound states at threshold since their tension is the modulus of an
algebraic sum rather than the arithmetic sum of moduli (positive
numbers).

\section{T-duality}\label{Tdual}

An alternative way to understand the geometry behind the
dy-charged string sectors relies on T-duality that transforms
pairs of {\sl obliquely} magnetized D9-branes into pairs of
intersecting unmagnetized D6-branes. Obviously the required
T-duality depends on the pair of branes under consideration and
for generic {\sl oblique} fluxes it is impossible to T-dualize the
complete set of magnetized D9-branes into a set of (neutral)
intersecting D6-branes. Yet, one can proceed pair by pair. The
procedure is particularly rewarding for ${\cal N}=2$ sectors where
the computation of $I_{ab}^\perp$ and the determination of the
`unmagnetized' directions ${u}_{ab}$ and thus of the generalized
KK momenta ${p}_{ab}$ carried by open strings
 is rather subtle if not ambiguous to some extent.

For definiteness let us consider two examples that illustrate the
general procedure: the sectors 5-8 and 5-4 of the AM
model\footnote{We use these subsectors only for illustrative
purposes.} \cite{am}.

In the first 5-8 case (in units of $1/4\pi^2\ap$) \be F_5 = - dx^1
dx^3 - dy^1 dy^3 \qquad F_8 = -dx^1 dy^1 + dx^2 dy^2 - dx^3 dy^3
\ee in the coordinate system where $x^i = x^i + 2\pi
k^i\sqrt{\ap}$, same for $y^i$. Barring the $(x^2, y^2)$ subtorus
where any T-duality does the job, there are two possible T-duality
trasformations of the remaining ${\bf T}^4$: $T_{y^1} T_{x^3}$ and
$T_{x^1} T_{y^3}$. Let us choose the first and combine it with
$T_{y^2}$ for definiteness. Neglecting the common non-compact spacetime
dimensions,
the problems is reduced to considering a ${\bf T}^6$
with two intersecting but unmagnetized D3-branes
spanning the worldvolumes \bea && D3_5 (a,b,c) = a (E_{x^1} + q_5
E_{\tx^3}) + b (E_{y^3} - q_5 E_{\ty^1}) + c E_{x^2}
\\
&& D3_8 (d,e,f) = d (E_{x^1} + q_8 E_{\ty^1}) + e (E_{y^3} - q_8
E_{\tx^3}) + f (E_{x^2} - q_8 E_{\ty^2}) \eea where $a,b,c,d,e,f$
are real parameters subject to periodic identifications
($a_i\approx a_i + 1$) and $E_{x^i}$ are orthogonal (since the
metric is diagonal at the susy point) vectors along the
(T-dualized) directions, normalized to \bea
&&|E_{x^1}| = r_1^2 = 2^{3/2} \quad  |E_{\ty^1}| = \tr_1^2 = 2^{-3/2} \\
&&|E_{x^2}|= r_2^2 = 2^{-1/2} \quad  |E_{\ty^2}|= \tr_2^2 = 2^{1/2} \\
&&|E_{\tx^3}| = \tr_3^2 = 2^{-1/2} \quad  |E_{y^3}|= r_3^2 =
2^{1/2} \quad . \eea Along $\tilde{{\bf T}}^2_{(2)}$ $D3_5$ and
$D3_8$ intersect once at an angle $\beta^{(2)}_{58}
=\arctan(\sqrt{2})$. Let us focus on the remaining $\tilde{{\bf
T}}^4$. We expect to find two orthogonal directions along which
the 5-8 strings carry momentum and winding. KK momentum $P_{58}$
lies along the common longitudinal direction \be P_{58} = \kappa {
E_{x^1} + q_8 E_{\ty^1} - q_5 q_8 E_{y^3}+q_5 E_{\tx^3} \over
|E_{x^1} + q_8 E_{\ty^1} - q_5 q_8 E_{y^3}+q_5 E_{\tx^3}|^2} \quad
, \ee that stretches once along the fundamental cell and has
length $|P_{58}|^2 = \kappa^2\sqrt{8}/15$ with $\kappa$ an
arbitrary integer. The allowed winding $W_{58}$ is aligned along
the unique direction which is orthogonal to both branes \be W_{58}
= \nu (E_{x^1} - 8 q_8 E_{\ty^1} - 2q_5 q_8 E_{y^3} - 4 q_5
E_{\tx^3}) \quad . \ee It winds $1 + 8 + 2 + 4 - (4-1) = 12$
times around the fundamental cell  of $T^4$ and has length
$|W_{58}|^2=\nu^2 15\sqrt{8}$. The minimal allowed value of $\nu$
is $1/15$.

We are left with the magnetized plane $\Pi_M$ spanned by the two
the vectors \bea && V_5 = E_{x^1} - 2 q_8 E_{\ty^1} + 2q_5 q_8
E_{y^3} + q_5 E_{\tx^3}
\\
&& V_8 = 2 E_{x^1} + 2 q_8 E_{\ty^1} + 3 q_5 q_8 E_{y^3} - 3 q_5
E_{\tx^3} \quad , \eea that lie along the worldline of the
projections of $D3_5$ and $D3_8$ in the two-plane orthogonal to
${\bf T}^2_{(2)}$, $P_{58}$ and $W_{58}$. The two vectors are such
that $|V_5|^2= 15/\sqrt{2}$, $|V_8|^2= 45/\sqrt{2}$ and $V_5 \cdot
V_8 = 15/\sqrt{2}$ and thus form an angle $\beta^{(1)}_{58} =
\arccos(1/\sqrt{3}) = \arctan(\sqrt{2})$! This being the same as
the one in the $T^2_{(2)}$ confirms that ${\cal N}=2$ susy is
preserved in this sector.

In order to compute $I^{(1')}_{58}$ one can T-dualize back and use
open / closed string duality that yields \be I_{ab}^\perp L_P
L_P^\prime = V(T^{tot}) \prod_{I\neq u} \sin(\pi\epsilon^I_{ab})
\sqrt{\det(1 +H_a)} \sqrt{\det(1 +H_b)} \ee Plugging numbers
$V(T^4)= 4$, $\sin(\pi\epsilon_{58}) = \sqrt{2/3}$, $L_P =
\sqrt{15/\sqrt{8}}$, $\sqrt{\det(1 +q_5 H_5)} = 5/4$,
$\sqrt{\det(1 +q_8H^\perp_8)}=3 \sqrt{3}/4$, $L_P^\prime = 15/L_W
= \sqrt{15/\sqrt{8}}$ one gets \be I_{58}^{(1')} = 1 \quad .\ee

Let us now turn our attention on the 4-5 sector of the AM model
\cite{am} \be F_5 = - dx^1\wedge dx^3 - dy^1 \wedge dy^3 \qquad
F_4 = - dx^2 \wedge dx^3 - dy^2\wedge dy^3 \ee A possible choice
for the T-duality transformation is $T_{y^1} T_{y^2} T_{x^3}$ that
yields \bea && D3_4 = a_4(E_{x^2} + q_4E_{\tx^3}) + b_4 (E_{y^3} -
q_4 E_{\ty^2}) + c_4 E_{x^1}
\\
&& D3_5 = a_5(E_{x^1} + q_5 E_{\tx^3}) + b_5 (E_{y^3} - q_5
E_{\ty^1}) + c_5 E_{x^2} \eea The allowed KK momenta lie along the
common longitudinal direction \be P_{45} = \kappa { q_4 E_{x^1} +
q_5 E_{x^2} + q_4 q_5 E_{\tx^3}  \over |q_4 E_{x^1} + q_5 E_{x^2}
+ q_4 q_5 E_{\tx^3} |^2} \ee that stretches once ($1+1+1 - (3-1)=
1$) along the fundamental cell of $T^6$ and has length $|P_{45}|^2
= \kappa^2 /3\sqrt{2}$ with $\kappa$ an integer. The allowed
windings stretch along the unique direction orthogonal to both
branes \be W_{45} = \nu (4q_5 E_{\ty^1} + q_4 E_{\ty^2} + E_{y^3})
\ee that winds $4 + 1 + 1 - (3-1) = 4$ times the fundamental cell,
the minimal allowed value of $\nu$ is $\nu = 1/6$ as can be seen
geometrically or by requiring $|W_{45}|_{min}^2 =
|P_{45}|_{min}^2$ for the minimal non-vanishing zero-modes.

\begin{figure}\label{Torox}
\includegraphics[width=.45\textwidth]{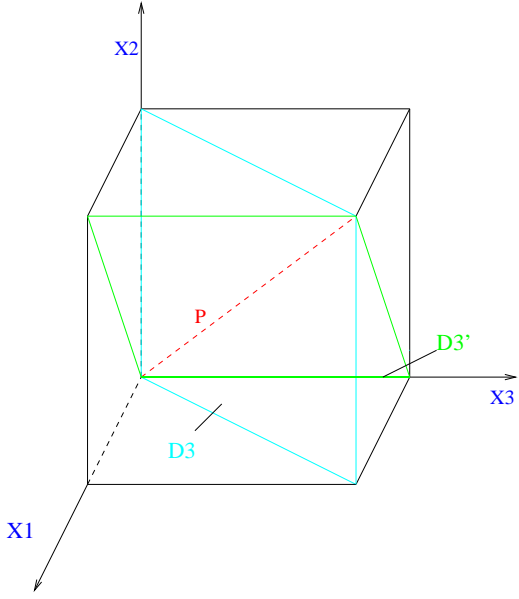}
\caption{D3$_5$ (D3) and D3$_4$ (D3') brane orientation wrt the fundamental cell of
$T^3_X$ is shown. The red dashed line along which the two planes intersect
indicate the unmagnetized direction.}
\end{figure}

The magnetized 4-plane $\Pi_M $ is spanned by the worldvolumes of
the two D2-branes \bea && D2_4^\perp = g_4 ( E_{x^1} - 2q_4q_5
E_{x^2} - 2q_5 E_{\tx^3}) + h_4 (E_{y^3} - q_4 E_{\ty^2} ) = g_4
\hat{E}_{g_4} + h_4 \hat{E}_{h_4}
\\
&& D2_5^\perp = g_5 ( E_{x^1} - 5 q_4q_5 E_{x^2} + q_5 E_{\tx^3})
+ h_5 ( E_{y^3} - q_5 E_{\ty^1}) = g_5 \hat{E}_{g_5} + h_5
\hat{E}_{h_5} \eea obtained neglecting the common longitudinal
direction $P_{45}$ \ie taking the orthogonal complements to
$P_{45}$ of $D3_4$ and $D3_5$. The hypervolumes spanned by
$\Pi_M$, $D2^\perp_4$, $D2^\perp_5$ are given by \be \widehat{V} =
\sqrt{\det (\hat{E}_A \cdot \hat{E}_B)} \ee where $\hat{E}_A =
\{\hat{E}_{g_4},\hat{E}_{h_4},\hat{E}_{g_5},\hat{E}_{h_5}\}$ for
$\Pi_M$, while $\hat{E}_a = \{\hat{E}_{g_4},\hat{E}_{h_4}\}$ for
$D2^\perp_4$ and $\hat{E}_b = \{\hat{E}_{g_5},\hat{E}_{h_5}\}$ for
$D2^\perp_5$. The relevant scalar products are \bea
&&|\hat{E}_{g_4}|^2 = 6\sqrt{2} \quad |\hat{E}_{h_4}|^2 =
2\sqrt{2} \quad |\hat{E}_{g_5}|^2 = 15\sqrt{2} \quad
|\hat{E}_{h_5}|^2 = {5\over 4} \sqrt{2} \\
&& \hat{E}_{g_4} \cdot \hat{E}_{g_5} = 6\sqrt{2} \quad
\hat{E}_{h_4} \cdot \hat{E}_{h_5} = \sqrt{2} \quad
 \hat{E}_{g_4} \cdot \hat{E}_{h_4} =  \hat{E}_{g_4} \cdot \hat{E}_{h_5} =
\hat{E}_{g_5} \cdot \hat{E}_{h_4} =  \hat{E}_{g_5} \cdot
\hat{E}_{h_5} = 0 \nonumber \eea So that \be V(\Pi_M) = 18 \quad
A(D2^\perp_4) = 2 \sqrt{6}  \quad A(D2^\perp_4) = {5\over 2}
\sqrt{6} \ee
\begin{figure}\label{Toroy}
\includegraphics[width=.5\textwidth]{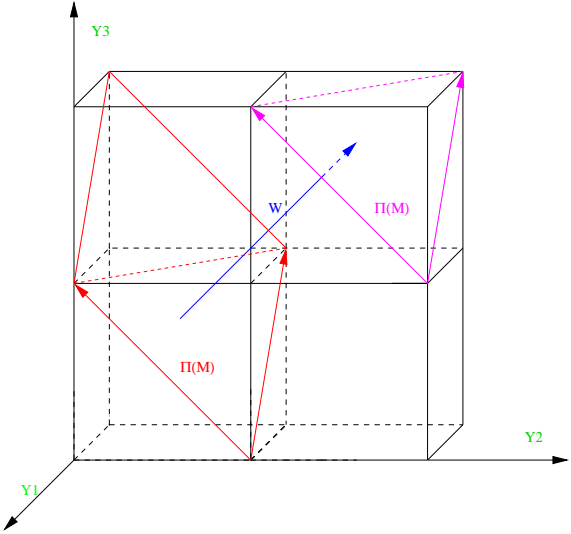}
\caption{The projection of $D3_4$ and $D3_5$  brane on $T^3_Y$
leads two D1 branes, along the direction $ \hat{E}_{h_4}$ and
$\hat{E}_{h_5}$ indicated with red arrows. They span the
magnetized plane $\Pi(M)$  orthogonal to the winding vector
$W_{45}$ (blue arrows).}
\end{figure}

Then the intersection angles are such that \be \prod_I
\sin(\pi\epsilon^I_{ab}) = V(\Pi_M)/ A(Dp_a) A(Dp_b) \ee In our
4-5 case, since  $\sin(\pi\epsilon^1_{45}) =
\sin(\pi\epsilon^2_{45})$ for susy reasons, one has \be
 \sin(\pi\epsilon_{45}) = \sqrt{18/30} = \sqrt{3/5}
\ee that means \be \tan(\pi\epsilon_{45}) =  \sqrt{3/2} \ee that
coincides with the result of diagonalizing $R_{45}$! In order to
compute the index $I_{45}^\perp$ and confirm that the minimal
winding is indeed $1/6$ it is very convenient to factorize the
problem in the two subtori ${\bf T}^6 = {\bf T}^3_{X} {\bf T}^3_{Y}$.
On the first
subtorus ${\bf T}^3_{X}= {\bf T}^3_{x^1x^2\tx^3}$, we have two D2-branes
intersecting along the common direction $P_{45}$ at an angle such
that $\cos(\beta^X_{ab})= \sqrt{2/5}$. The fig.(1) displays the
the orientation of $P_{45}$ wrt the  cell.
 As already stated $P_{45}$
winds only once within the fundamental cell. In the second
sub-torus ${\bf T}^3_{X}= {\bf T}^3_{\ty^1\ty^2 y^3}$ the two D1-branes
interesect only at the origin and span a plane orthogonal to
$W_{45}$, as fig.(2) shows.

The distance between two such planes, \ie intersections,
along the direction of $W_{45}$ is $|W_{45}|/6$ hence $\nu_{min} =
1/6$. Thus $I^\perp_{45} =1$, which is nicely consistent with open -
closed channel duality after putting numbers $V(T^6) = \sqrt{8}$,
$\sin(\beta) = \sqrt{3/5}$, $\sqrt{\det(1+H_4)}= 2$ and
$\sqrt{\det(1+H_5)}= 5/4$, $L_P= \sqrt{3\sqrt{2}}$ and $L_P^\prime
= 6/L_W = \sqrt{6/\sqrt{2}}$.

\section{Gauge Coupling Thresholds}\label{thresh}

Once the perturbative open string spectrum  is known, computing
some of the low-energy effective couplings is quite
straightforward. Tree level gauge couplings were determined in
\cite{mbet} and turned out to be given by \be {4\pi \over g_a^2} =
e^{-\Phi} \sqrt{\det({\cal G}_a + {\cal F}_a)} = e^{-\Phi} |W_a|
\sqrt{\det({G} + {F}_a)} \ee Moduli dependence is hidden inside
the induced metric ${\cal G}_a$. The induced internal magnetic
field ${\cal F}_a= F^a_{ij} \ \partial_\alpha X^i \partial_\beta
X^j $ satisfies the standard Dirac quantization condition. At
supersymmetric points, the expression simplifies and indeed
coincides with the WZ term due to the identity of tension and
charge for the magnetized brane (configuration). Moreover, in
principle, for proper choices of the {\sl oblique} magnetic fields
all closed string moduli, except for the complexified dilaton,
could be frozen. The ratios of the couplings would then be
completely determined. More precisely, we are ignoring possible
mixings of the open string Wilson line moduli, \ie we are setting
their VEV's to zero for the time being.

Although a consistent variant of the AM model \cite{am} has not
yet been found, our aim  is to extend the tree level analysis
\cite{mbt} to one-loop and derive the running of the gauge
couplings as well as their threshold corrections. This is a
preliminary step towards the study of gauge coupling unification
in models with {\sl oblique} magnetic fluxes or closely related
(orbifold) models that might be phenomenologically more appealing
but still solvable.

We will follow the strategy pioneered by \cite{bp,bachfabre} and
successfully applied to type I orbifolds in
\cite{abd,mbjfmrgflow}, to generic  type I vacuum configurations
in \cite{mbjfmrgflow} and to intersecting brane models in
\cite{stieb}, based on the background field method.

As hinted at in the introduction, the  method consists in applying
an abelian, constant and small magnetic field in some spacetime
directions, computing the effect of such an integrable deformation
and then extracting the desired (quadratic) term in the one-loop
effective action.

Only open strings that have at least one end on the spacetime
magnetized brane will sense the presence of the magnetic field and
can a priori contribute to the renormalization of  the
corresponding gauge coupling. In principle, one should consider
dipole strings, preserving ${\cal N}=4 $, as well as singly- and
doubly-charged strings, preserving ${\cal N}=1$ or ${\cal N}=2$.
However, similar to what happens in simpler cases with parallel
magnetic fields or untwisted sector of orbifolds, ${\cal N}=4$
sectors neither contribute to the running nor to the thresholds,
while ${\cal N}=1$ or ${\cal N}=2$ sectors
 contribute both to the running and
to the thresholds. Massless open string states  contribute to the
logarithmic running, and we will retrieve the field-theory
$\beta$-function coefficients studying the IR limit of the
relevant one loop amplitudes. The contribution to the thresholds
from ${\cal N}=2$ sectors is particularly simple since the gauge
coupling is 1/2 BPS-saturated
\cite{lerche,bachfabre,abd,mbjfmrgflow}, only the zero-modes coded
in the `magnetically' deformed internal lattice sum will survive
but no string oscillator modes \cite{eliasrev}. The contribution
to the thresholds from ${\cal N}=1$ sectors is slightly more
involved since the gauge coupling is not BPS-saturated in this
case \cite{abd,mbjfmrgflow}. As obvious from the discussion of the
spectrum in Section \ref{partfunct} there are no lattice sums in
these sectors, but magnetically shifted string oscillator modes
can and do contribute. Luckily we will recast our anlysis along
the lines of \cite{stieb} where the modular integral for the case
of intersecting branes were computed and finally expressed in
terms of $\Gamma$ functions. Once again the moduli dependence
hidden in the lattice sums or the magnetic shifts (T-dual to
angles) can be fixed for particular choices of the internal {\sl
oblique} fluxes.

\subsection{General analysis}

For definiteness, we turn on an abelian magnetic field in
spacetime directions 2 and 3, {\it viz.} \be F_{\m\n} =
\delta_{[\m}^2 \delta_{\n]}^3 f Q \ee where $Q$ is one of the
generator  of the unbroken CP group, normalized so that
$Tr_N(Q)=0$ and $Tr_N(Q^2)=1/2$. Depending on the embedding of $Q$
in the CP group one can find different behaviours. We will mostly
focus on the case in which $Q$ is a generator of a non-abelian and
thus non-anomalous factor.

As for internal fluxes, the spacetime  magnetic deformation is
integrable. Amplitudes on surfaces with no boundaries, such as
torus and the Klein Bottle are insensible to the external field.
Annulus and M\"obius strip do couple to the external field and the
connected generating functional depends on $f$. The main effects
of turning on $f$ are the magnetic shifts $\epsilon^Q_{ab}$ of the
transverse spacetime modes and the degeneracy $I^Q_{ab}$ of Landau
levels for the string modes in the $[23]$ plane. Both are related
to the charge $Q$ of the open string according to\footnote{To be
pedantic for the annulus $Q_a$ actually means $Q_{(a)}\otimes
1_{(b)}$ and $Q_a + Q_b$ actually means $Q_{(a)}\otimes 1_{(b)}+
1_{(a)}\otimes Q_{(b)}$.} \be
 \epsilon^Q_{ab} =  {1\over \pi} [\arctan(Q_a f) +\arctan(Q_bf)]
\ee
for ${\cal A}_{ab}(f)$ with $I^Q_{ab} = (Q_a + Q_b) f/2\pi $, and
\be
\epsilon^Q_{aa} =  {2\over \pi} \arctan(Q_a f)
\ee
for ${\cal M}_{a a}(f)$ with $\hat{I}^Q_{a a} = 2 Q_a f/2\pi$.

Expanding the Annulus and the M\"obius amplitudes up to second
order in $f$  one gets the one-loop gauge threshold for the group
$Q$ belongs to \cite{abd,bachfabre} \be \Delta_Q  =  \int
\frac{dt}{t}({\cal A}_Q''(0) + {\cal M}_Q''(0)) = \int
\frac{dt}{4t} B_Q(t) \ee that implicitly depends on the moduli
fields  through the dependence on the latter of the masses of the
unoriented open string states running in the loop. We use $Q$ to
label the (factor) group we are computing the threshold of.

The only allowed momenta are along the light-cone  directions $0$
and $1$. Analytic continuation and volume regularization thus
yield \be \int {V_{LC} dp^+ dp^- \over (2\pi)^2} \exp(-\pi \ap t
p^+p^-) = {V_{LC}
 \over (2\pi)^2 \ap t } \quad .\ee One can easily write down
the contribution of singly as well as doubly charged unoriented
strings\footnote{Sums over $a$ or $b$  include branes as well as
their images under $\Omega$, which in our conventions sends $q$
into $-q$.}: \bea {\cal A}^{{\cal N}=1}_{Q}(f) & = & \sum_{a,b}
I_{ab} \sum_{\a,\b} {c_{\a\b} \over 2} tr_{N_a\times N_b} \left[
{(Q_a + Q_b)f \over (2\pi)^3 \ap t}
{\theta\left[^\a_\b\right](\epsilon_{ab}^{Q} \tau_A |\tau_A) \over
\theta_1(\epsilon_{ab}^{Q} \tau_A| \tau_A)}\right]
\prod_I{\theta\left[^\a_\b\right](\epsilon^I_{ab} \tau_A|\tau_A)
\over
\theta_1(\epsilon^I_{ab} \tau_A| \tau_A)} \nonumber\\
{\cal M}^{{\cal N}=1}_{Q} (f)& = & -  \sum_{a}\hat{I}_{a
a}\sum_{\a\b}c_{\a\b}\; Tr_{N_a} \left[\frac{Q_a f}{(2\pi)^3 \a t
}\frac{\theta[^\a_\b](\epsilon^Q_{a a} \tau_A|\tau_M)}
{\theta_1(\epsilon^Q_{a a}\tau_A|\tau_M)}\right]\prod_I
\frac{\theta[^\a_\b](\epsilon^I_{a a}\tau_A|\tau_M)}
{\theta_1(\epsilon^I_{a a}\tau_A|\tau_M)}) \eea where $\tau_A =
it/2$, $\tau_M = \tau_A + 1/2$, and, denoting as usual by $u$ the
internal unmagnetized direction, when present, \bea \quad {\cal
A}^{{\cal N}=2}_{Q}(f) & = &  -i \sum_{a,b}\
I^\perp_{ab}\Lambda^u_{ab}(\tau_A) \sum_{\a,\b} {c_{\a\b} \over 2}
tr_{N_a\times N_b}\left[ {(Q_a + Q_b)f \over (2\pi)^3 \ap t}
{\theta\left[^\a_\b\right](\epsilon_{ab}^{Q} \tau_A |\tau_A) \over
\theta_1(\epsilon_{ab}^{Q}\tau_A|\tau_A)}\right]\nonumber\\
&  & {\theta\left[^\a_\b\right](0|\tau_A) \over \eta^3(\tau_A)}
 \prod_{I\neq u}{\theta\left[^\a_\b\right] (\epsilon^I_{ab}\tau_A|\tau_A)
\over \theta_1( \epsilon^I_{ab}\tau_A|\tau_A)} \eea \bea {\cal
M}^{{\cal N}=2}_{Q}(f) & = & i\sum_{a}\hat{I}_{a a}^{\perp}
\Lambda^{u}_{a a}(\tau_A) \sum_{\a\b}c_{\a\b}N_a Tr_{N_a}
\left[\frac{Q_a f}{(2\pi)^3 \a t
}\frac{\theta[^\a_\b](\epsilon^Q_{a a} \tau_A|\tau_M)}
{\theta_1(\epsilon^Q_{a a}\tau_A|\tau_M)}\right]
\frac{\theta[^\a_\b](0|\tau_M)}{\eta^3(\tau_M)}\nonumber\\
& & \prod_{I\neq u} \frac{\theta[^\a_\b](\epsilon^I_{a
a}\tau_A|\tau_M)} {\theta_1(\epsilon^I_{a a}\tau_A|\tau_M)} \eea

${\cal N}=4$ sectors, such as neutral  and dipole strings with
opposite $Q$ charge at their ends do not contribute to the
thresholds since their modes are not shifted and they simply
receive an overall factor reflecting  the 'magnetic deformation'
of the lattice sum. Similarly, the M\"obius strip does not
contribute if $Q$ is part of $SO(N_0)$, associated to
`unmagnetized' branes if present, so that the two ends have
opposite charge. Henceforth we set $\ap = 1/2$ for convenience.

\subsection{${\cal N}=1$ sectors}

In ${\cal N}=1$ supersymmetric sectors,  expanding to quadratic
order one gets \bea {\cal A}^{{\cal N}=1}_{Q} (f) & = &
i\sum_{a,b}\ I_{ab}\frac{1}{2} \left(\frac{f}{2\pi}\right)^2
\sum_{\a\b}\frac{c_{\a\b}}{{8\pi^2}}\; Tr_{N_a\times N_b}(Q_a
+Q_b)^2\frac{\theta''[^\a_\b](0)}{\eta^3} \prod_I
\frac{\theta[^\a_\b](\epsilon^I_{ab}\tau_A|\tau_A)}
{\theta_1(\epsilon^I_{ab}\tau_A|\tau_A)}) + ... \nonumber\\
{\cal M}^{{\cal N}=1}_{Q} (f) & = &  -i\sum_{a}\ I_{a
a}\frac{1}{2} \left(\frac{f}{2\pi}\right)^2
\sum_{\a\b}\frac{c_{\a\b}}{2\pi^2}\; Tr_{N_a}
(Q_a^2)\frac{\theta''[^\a_\b](0|\tau_M)}{\eta^3(\tau_M)} \prod_I
\frac{\theta[^\a_\b](\epsilon^I_{a a}\tau_A|\tau_M)}
{\theta_1(\epsilon^I_{aa} \tau_A|\tau_M)} + ... \eea Summing over
spin structures and using the generalized  Jacobi $\theta$
function identity \be \sum_{\a\b}
c_{\a\b}\frac{\theta''[^\a_\b](0)}{\eta^3} \prod_I
\frac{\theta[^\a_\b](\epsilon^I\tau)} {\theta_1(\epsilon^I\tau)} =
2\pi \sum_I \frac{\theta_1'(\epsilon^I_{ab}\tau)}
{\theta_1(\epsilon^I\tau)}  \ee give \bea\label{Ba} B_{Q}^{{\cal
N}=1}(t)  & = & \frac{i}{\pi} \sum_{a,b}  I_{ab} Tr_{N_a\times
N_b}(Q_a +Q_b)^2 \sum_I
\frac{\theta_1'(\epsilon^I_{ab}\tau_A|\tau_A)}
{\theta_1(\epsilon^I_{ab}\tau_A|\tau_A)} \nonumber\\
\hat{B}_{Q}^{{\cal N}=1}(t)  & = & -\frac{i}{\pi} \sum_{a} I_{aa}
Tr_{N_a}(2 Q_a^2) \sum_I \frac{\theta_1'(\epsilon^I_{a
a}\tau_A|\tau_M)} {\theta_1(\epsilon^I_{a a}\tau_A|\tau_M)} \eea
At this point it is easy to extract the $\b$-function coefficents
from the IR limit of (\ref{Ba}) \bea b_{Q}^{{\cal N}=1} & = &
\frac{1}{2}\sum_{a,b} I_{ab}
Tr_{N_a\times N_b}(Q_a +Q_b)^2  \nonumber\\
\hat{b}_{Q}^{{\cal N}=1} & = & - \frac{1}{2} \sum_{a} I_{a a}
Tr_{N_a}(2Q_a)^2\eea Since all vectors belong to ${\cal N}=4$
multiplets, $\b$-function are positive, \ie all non-abelian
couplings grow in the UV.

In order to perform the integral and  compute $\Delta_Q$ we switch
to the transverse channel and end up with the following
expressions \bea\label{thr} \Delta_{Q}^{{\cal N}=1} & = &
\frac{1}{\pi}\sum_{a,b} I_{ab} Tr_{N_a\times N_b}(Q_a +Q_b)^2
\sum_I\int_{0}^{\infty}\frac{\theta_1'(\epsilon^I_{ab}|i\ell)}
{\theta_1(\epsilon^I_{ab}|i\ell)}\;d\ell \nonumber\\
\hat\Delta_{Q}^{{\cal N}=1} & = & -\frac{1}{2\pi} \sum_a
Tr_{N_a}(2Q_a)^2 I_{a a}
\sum_I\int_{0}^{\infty}\frac{\theta_1'(\epsilon^I_{0a}|i\ell +
1/2)} {\theta_1(\epsilon^I_{0a}|i\ell + 1/2)}\;d\ell \eea

Series expansion \be
\frac{{\theta_1'(\epsilon|\tau)}}{\theta_1(\epsilon|\tau)} = \pi
\cot(\pi\epsilon) + 2 \sum_{k=1}^\infty \zeta(2k) \epsilon^k
(E_{2k}(\tau) - 1) \quad ,\ee where $ \zeta(2k) =
(2\pi)^{2k}|B_{2k}| / (2k)!$ and $E_{2k}(\tau)$ is an Eisenstein
series with modular weight $2k$, expose the potentially divergent
terms \bea \delta_{Q}^{{\cal N}=1} & = &
\sum_{a,b}\frac{I_{ab}}{4}Tr_{N_a\times N_b}(Q_a +Q_b)^2
 \sum_I\cot(\pi \epsilon^I_{ab}) \int_0^\infty d\ell
\nonumber\\
\hat\delta_{Q}^{{\cal N}=1} & =  & -2 \sum_{a}\hat{I}_{aa}
Tr_{N_a}(Q_a^2)
 \sum_I\cot(\pi \epsilon^I_{0a}) \int_0^\infty d\ell
\eea that eventually cancel thanks to (NS-NS)  tadpole
cancellation, for the non-anomalous $Q$, with $Tr(Q)=0$. The
latter condition has to be used in order to dispose of the
divergent terms with $f$ insertions in two different boundaries.
Divergences from insertions on the same boundary cancel between
annulus and M\"obius strip thanks to tadpole cancellation.

The finite terms boil down to integrals of the form \cite{stieb}
\be \int_0^\infty d\ell \sum_{k=1}^\infty 2\zeta(2k) \epsilon^k
(E_{2k}(i\ell) - 1) = - \pi\log\left[\Gamma(1 - \epsilon) \over
\Gamma(1 + \epsilon)\right] + 2\pi\epsilon \gamma_E \ee \be
\int_0^\infty d\ell  \sum_k 2\zeta(2k) \epsilon^k (E_{2k}(i\ell +
1/2) - 1) = - \pi\log\left[\Gamma(1 - 2\epsilon) \over \Gamma(1 +
2\epsilon)\right] + 2\pi\epsilon \gamma_E \ee Actually the last
contributions, linear in $\epsilon$, drop after summing over the
three internal directions in supersymmetric cases.

Summing the various contributions one finally gets
\bea\Delta^{{\cal N}=1}_{Q} & = & - \sum_{a,b}
\frac{I_{ab}}{2}Tr_{N_a\times N_b}(Q_a^2 +Q_b^2) \sum_I
\log\left[\Gamma(1 - \epsilon^I_{ab}) \over \Gamma(1 +
\epsilon^I_{ab})\right]
\nonumber\\
\hat\Delta^{{\cal N}=1}_{Q } & = & \sum_{a} I_{a a} Tr_{N_a}(2
Q_a)^2\sum_I \log\left[\Gamma(1 - \epsilon^I_{aa}) \over \Gamma(1
+ \epsilon^I_{aa})\right]  \quad ,\eea where $\epsilon^I_{aa} = 2
\epsilon^I_{ao}$.

\subsection {${\cal N}=2$ sectors}

Thresholds corrections from ${\cal N}=2$ sectors  are much easier
to compute since they correspond to BPS saturated couplings.
Indeed, for ${\cal N}=2$ supersymmetric sectors, the terms
quadratic in $f$ read \bea {\cal A}^{{\cal N}=2}_{Q}(f) & = &
\sum_{a,b} I_{ab}^{\perp}\Lambda^{u}_{ab}(\tau_A)
\frac{1}{2}\left(\frac{f}{2\pi}\right)^2
\sum_{\a\b}\frac{c_{\a\b}}{8\pi^2}\; Tr_{N_a\times N_b}(Q_a
+Q_b)^2 \frac{\theta''[^\a_\b](0|\tau_A)}{\eta^3(\tau_A)}
\frac{\theta[^\a_\b](0|\tau_A)}{\eta^3(\tau_A)} \nonumber\\
&  & \prod_{I\neq u}
\frac{\theta[^\a_\b](\epsilon^I_{ab}\tau_A|\tau_A)}
{\theta_1(\epsilon^I_{ab}\tau_A|\tau_A)} + ... \nonumber\\
{\cal M}^{{\cal N}=2}_{Q} (f) & = &  \sum_a I_{a a}^{\perp}
\Lambda^{u}_{aa}(\tau_A) \frac{1}{2}\left(\frac{f}{2\pi} \right)^2
\sum_{\a\b}{c_{\a\b}\over 8 \pi^2}\; Tr_{N_a}(2Q_a)^2
\frac{\theta''[^\a_\b](0|\tau_M)}{\eta^3(\tau_M)}\nonumber\\
&  & \frac{\theta[^\a_\b](0|\tau_M)}{\eta^3(\tau_M)} \prod_{I\neq
u} \frac{\theta[^\a_\b](\epsilon^I_{a a}\tau_A|\tau_M)}
{\theta_1(\epsilon^I_{a a}\tau_A|\tau_M)}) + ... \eea and the
Jacobi $\theta$ function identity \be \sum_{\a\b}
c_{\a\b}\frac{\theta''[^\a_\b](0)}{\eta^3}
\frac{\theta[^\a_\b](0)}{\eta^3} \prod_{I\neq u}
\frac{\theta[^\a_\b](\zeta^I|\tau )} {\theta_1(\zeta^I|\tau )} =
4\pi^2 \quad , \ee valid for $\sum_I \zeta^I =0$, imply that only
the lattice sum over 1/2 BPS states contributes.

Manipulations similar to the above yield the following results for
$\b$-function  coefficents in ${\cal N}=2$ sectors, \bea
b_{Q}^{{\cal N}=2} & = & \sum_{a,b}I_{ab}^{\perp} Tr_{N_a\times
N_b}(Q_a +Q_b)^2
\nonumber\\
\hat{b}_Q^{{\cal N}=2} & = & - \sum_{a} \hat{I}_{a a}^{\perp}N_a
Tr_{N_a}(2 Q_a)^2  \eea Since all vectors belong to ${\cal N}=4$
multiplets,  $\b$-functions are positive, \ie all non-abelian
couplings grow in the UV.

For the thresholds one has \bea \Delta_{Q}^{{\cal N}=2} & = &
\frac{1}{2} \sum_{a,b}I_{ab}^{\perp} Tr_{N_a\times N_b}(Q_a
+Q_b)^2\int_{0}^{\infty}\Lambda_{ab}^u(it)\; \frac{dt}{t}
\nonumber\\
\Delta_{Q}^{{\cal N}=2} & = & - \frac{1}{2} \sum_{a} I_{a
a}^{\perp} Tr_{N_a}(2 Q_a)^2\int_{0}^{\infty}\Lambda_{a a}^u(it)\;
\frac{dt}{t} \eea The integrals of the `regulated' lattice sums
can be performed with the aid of the formula \be\label{int}
\int_0^\infty \frac{dt}{t} \sum_{(k_1,k_2)\neq (0,0)}
\exp(-\pi{\ell}  |k_1 + U k_2|^2/V_2 U_2) = \gamma_E - \log[4\pi
V_2 U_2 |\eta(U)|^4] \ee where $V_2$ is the volume and $U$ the
complex structure of the  unmagnetized torus. Inserting into
(\ref{int}) one gets \bea \Delta_{Q}^{{\cal N}=2} & = &
\frac{1}{2}\sum_{a,b} I_{ab}^{\perp}Tr_{N_a\times N_b} (Q_a
+Q_b)^2[
\ln(V_2 U_2|\eta(U)|^4 4\pi) -\gamma_E]\nonumber\\
\Delta_{Q}^{{\cal N}=2} & = & -2 \sum_a I_{aa}^{\perp}
Tr_{N_a}(Q_a^2) [ \ln(4 V_2 U_2|\eta(U)|^4 4\pi) - \gamma_E] \eea

For proper choices of the internal {\sl oblique} fluxes all closed
string moduli are fixed, modulo mixing with massless  open string
states, and the above formulae give simply numbers as we will
momentarily see.

We would now like to comment on the effect of large extra
dimensions on the coupling costants running. Setting $U_1=0$ for
simplicity and expanding the logarithm in the last expression
yields \be \ln(4\pi V_2 U_2|\eta(U)|^4)= \ln(4\pi V U_2)-\frac{\pi
r_2}{3r_1} + 4\sum_n\ln (1-e^{-2\pi\frac{r_2}{r_1}n})\ee where
$r_2\,r_1$ are the radii of $T^2_{u}$ so that $U_2 = r_2/r_1$. One
can envisage two different situations
\begin{itemize}
\item $r_2\sim r_1$, when the radii are fixed at the string scale,
and corrections do not  seem to affect the usual logarithmic
behavior.

\item $r_2>>r_1$. Power corrections  are dominant, this is the
scenario already described in \cite{DudGher}, where power
corrections are induced by the running in the loop of bulk
particle, with KK towers organized in ${\cal N}=2$ multiplets.
Power law behavior can be exploited to lower the unification
scale. This is achieved  if one of the unmagnetized eigenvectors
points toward a large extra dimension.
\end{itemize}

\section{Outlook}\label{fine}

In the present paper we have derived  explicit formulae for the
one-loop contributions to type I string compactifications on tori
with arbitrary magnetic fluxes \cite{am,akm, mbet}. We have
checked consistency with the transverse channel and identified the
correct tadpole conditions. Further insights in the geometry of
these vacuum configurations has been gained by means of T-duality
\cite{bala,berk}. We have then turned our attention to the
one-loop threshold corrections to the non-abelian gauge couplings
and derived very compact expressions thereof, relying on similar
analyses for type I orbifolds \cite{abd,mbjfmrgflow} and
intersecting brane models \cite{stieb}. Although unrealistic in
many respects, toroidal models of this kind may be used as
building blocks or rather starting points for type I orbifolds and
other solvable (supersymmetric) compactifications. In particular
the emergence of induced lower dimensional R-R charges and NS-NS
tensions of both signs plays a crucial role in solving
\cite{marche,dt} some long standing puzzles \cite{aspuzzle}.

Given the high level of control one has on this class of models,
one can restrict one's attention  onto those that resemble as
closely as possible the Standard Model or some of its
supersymmetric or grand unified generalizations. In principle, the
magnetic fields can be tuned so as to produce the desired gauge
group and fermionic content, and achieve gauge coupling
unification.  With more effort one can try to generate the correct
pattern of Yukawa couplings and trigger supersymmetry breaking in
a controllable way \cite{susybreak}.

Stabilizing dilaton and axion   may require switching to a T-dual
description in terms of D3-branes that allows the introduction of
closed string 3-form fluxes. Yet the same goal may be achieved by
means of non-perturbative effects such as D5-brane and D-string
instantons. In any case, at present the possibility that all
moduli be stabilized by perturbative effects remains a challenge.
The presence of dilaton tadpoles at different orders in
perturbation theory may help achieving this goal.

Moreover, open string Wilson line moduli,  especially those
charged under the anomalous $U(1)$'s, can mix with closed string
moduli, due to their contribution to D-terms\footnote{We thank L.
Ibanez and the referee of \cite{mbet} for vigorously pointing this
out to us.}. This complicates the analysis, that has been so far
performed at the origin of the open string moduli space. In this
respect, it is reassuring to observe that scalars in vector
multiplets can be lifted by orbifold projections and in any case
they can be treated exactly along the lines of \cite{bps}. In
order to set the stage for the discussion of the lifting of
scalars in chiral multiplets one should compute the
superpotential, \ie the Yukawa couplings \cite{yukawa}. For
charged open string states, the relevant amplitudes involve
mutually non-abelian twists in general and the perspective of
computing them is daunting\footnote{As suggested by C. Bachas and
E. Kiritsis it may prove convenient to extract the coupling from
factorization of a four-point amplitude.}. Yet it may be worth
proving.

\section*{Acknowledgements}
We would like to thank P.~Anastasopoulos, C.~Angelantonj,
I.~Antoniadis, S.~Ferrara, F.~Fucito, A.~Lionetto, J.~F.~Morales
Morera, G.~Pradisi, M.~Prisco,  A.~Sagnotti, and Ya.~Stanev for
useful discussions. During completion of this work M.B. was
visiting the String Theory group at Ecole Polytechnique,
E.~Kiritsis and his colleagues are warmly acknowledged for their
kind hospitality.  This work was supported in part by INFN, by the
MIUR-COFIN contract 2003-023852, by the EU contracts
MRTN-CT-2004-503369 and MRTN-CT-2004-512194, by the INTAS contract
03-516346 and by the NATO grant PST.CLG.978785.
\newpage
\section*{Appendix A: Some useful formulae}

\bea
\prod_{I\neq u} \sin(\pi \epsilon^I_{ab}) & = &
\prod_{I\neq u} {1\over 2i}(e^{i\pi \epsilon^I_{ab}}-
e^{-i\pi \epsilon^I_{ab}})\nonumber\\
& = & \sqrt{\prod_{I\neq u} e^{-i\pi \epsilon^I_{ab}}\left({1\over
2i} \right) (e^{i 2\pi \epsilon^I_{ab}} - 1) e^{i\pi
\epsilon^I_{ab}} \left(- {1\over 2i} \right)
(e^{-i 2\pi \epsilon^I_{ab}} - 1)}  \nonumber\\
& = &\sqrt{\det'\left({R_{ab} - 1\over 2} \right)} =
{\sqrt{\det'(H_a + H_b)} \over \sqrt{\det(1 + H_a)} \sqrt{\det(1 +
H_b)}} \eea

\bea
\prod_{I\neq u} \cos(\pi \epsilon^I_{ab}) & = &
\prod_{I\neq u} {1\over 2}(e^{i\pi \epsilon^I_{ab}}+
e^{-i\pi \epsilon^I_{ab}}) \nonumber\\
& = & \sqrt{\prod_{I\neq u} e^{-i\pi \epsilon^I_{ab}}\left({1\over
2} \right) (e^{i 2\pi \epsilon^I_{ab}} + 1) e^{i\pi
\epsilon^I_{ab}} \left({1\over 2} \right)
(e^{-i 2\pi \epsilon^I_{ab}} + 1)}  \nonumber\\
& = &\sqrt{\det\left({R_{ab} + 1\over 2} \right)} = {\sqrt{\det(1
+ H_a H_b)} \over \sqrt{\det(1 + H_a)} \sqrt{\det(1 + H_b)}}
\nonumber\\ & = & Tr(U_{ab}) = Tr(U_{a}U_{b}) \eea where $U_a = {
{\rm AExp}(-{1\over 2}F_{a,ij}\Gamma^{ij}) \over \sqrt{\det(1 +
H_a)}}$. As a corollary \be \prod_{I\neq u} 2 \cos(\pi
\epsilon^I_{a}) = {1 \over \sqrt{\det(1 + H_a)} } = Tr_s(U_{a})
\ee
\be
 2 \sum_{I\neq u} \cos(2\pi \epsilon^I_{ab}) =  \sum_{I\neq u}
(e^{2i\pi \epsilon^I_{ab}}+ e^{-2i\pi \epsilon^I_{ab}}) =
Tr_v(R_{ab}) = Tr_v(R_{a}R_{b})
\ee
\section*{Appendix B: Theta functions}
\subsection*{Definitions}
In order to fix notations, we report in this appendix the Jacobi
$\theta$-functions, we used
throughout the paper. Let $q=e^{2\pi i \tau}\,$ they are defined as guassian
sums
\be
\theta\left[^\a_\b\right](z|\tau) =\sum_{n} q^{{1\over 2} (n -\a)^2}
e^{2\pi i (z-\b)(n-\a)}
\ee
where $\a \ \b \in R$.\\
Equivalently, for  particular values of characteristics, such as
$\,\a\; \b = 0\mbox{ , }\frac{1}{2}\;$  they are  given also
in terms of infinite product as follows
\bea
\theta\left[^\frac{1}{2}_\frac{1}{2}\right](z|\tau) = \theta_1(z|\tau)
& = & 2 q^{1\over 8}\sin(\pi z)
\prod_{m=1}^{\infty}(1-q^m)(1-e^{2\pi i z}q^m)(1-e^{-2\pi i z}q^m)\nonumber\\
\theta\left[^\frac{1}{2}_0\right](z|\tau) = \theta_2(z|\tau)
& = & 2 q^{1\over 8}\cos(\pi z)
\prod_{m=1}^{\infty}(1-q^m)(1+e^{2\pi i z}q^m)(1+e^{-2\pi i z}q^m)\nonumber\\
\theta\left[^0_0\right](z|\tau) = \theta_3(z|\tau)
& = & \prod_{m=1}^{\infty}(1-q^m)
(1+e^{2\pi i z}q^{m-\frac{1}{2}})(1+e^{-2\pi i z}q^{m-\frac{1}{2}})\nonumber\\
\theta\left[^0_\frac{1}{2}\right](z|\tau) = \theta_4(z|\tau)
& = & \prod_{m=1}^{\infty}(1-q^m)
(1-e^{2\pi i z}q^{m-\frac{1}{2}})(1-e^{-2\pi i z}q^{m-\frac{1}{2}})
\eea
The Dedekind function $ \eta$ is defined as
\be
\eta(\tau) = q^{1\over 24}\prod_{n=1}^{\infty} (1-q^n)
\ee
The Eisenstein series are
\be
E_r = \sum_{m\neq 0}^{\infty}\sum_{n \neq 0}^{\infty} \frac{1}{(m + n\tau)^r}
\ee
with $r > 2 $.
Moreover they can be expressed  as polynomial of elliptic funtions
\be
E_{2k}(\tau) = 1 + {(2\pi i)^{2k}\over (2k-1)!\zeta(2k)}
\sum_{n=1}^\infty \sigma_{2k-1}(n) e^{2\pi i n \tau} \ee where  $\zeta(2k)$
is Riemann zeta function and $\sigma_{2k-1}(n)$ is the divisor function
\be
\sigma_{2k-1}(n) = \sum_{d|n} d^{2k-1} \quad.
\ee
\subsection*{Modular Transformations}
Under T and S modular trasformations on the arguments the functions, given
above, have pecular properties:
\bea
\theta\left[^\a_\b\right](z|\tau +1) & = & e^{-i\pi\a(\a-1)}\
\theta\left[^\a_{\b+\a-\frac{1}{2}}\right](z|\tau) \nonumber\\
\eta(\tau + 1) & = & e^{{i\pi\over12}}\,\eta(\tau) \nonumber\\
E_{2k}(\tau + 1 ) & = &  E_{2k}(\tau) \nonumber\\
\theta\left[^\a_\b\right](\frac{z}{\tau}|-{1\over \tau} ) & = & (-i\tau)^
{1\over2} \,e^{2i\pi\a\b +i\pi z^2/\tau}\ \theta\left[^\b_{-\a}\right](z|\tau)
\nonumber\\
\eta(-{1\over\tau}) & = & (-i\tau)^{1\over 2} \,\eta(\tau)\nonumber\\
E_{2k}(-{1\over\tau}) & = & \tau^2k E_{2k}(\tau)
\eea
The modular transformation P on the Jacobi functions is more involved
as it consists in a sequence of T and S transformation ($P=TST^2S$).
on the modular parameter  $\tau_M = \frac{1}{2} + \frac{it}{2}$
\bea
\theta\left[^\a_\b\right]({z\over i t}|\frac{1}{2} + \frac{i}{2t}) & = &
e^{-i\pi\a(\a-1) - 2\pi i (\a+\b-1/2)^2 +2\pi z^2/t}\, \sqrt{-it}\;
\theta\left[^{\a+2\b-2}_{1/2 -\a-\b}\right](z|\frac{1}{2} + \frac{it}{2})
\nonumber\\
\eta(\frac{1}{2} + \frac{i}{2t}) & = & e^{i\pi/4} \sqrt{-it}\;
\eta(\frac{1}{2} + \frac{it}{2})
\eea

\end{document}